# Kinetic study of the CN + $C_2H_6$ hydrogen abstraction reaction based on an analytical potential energy surface

Joaquin Espinosa-Garcia*,[a] and Somnath Bhowmick*,[b]

a) Departamento de Química Física and Instituto de Computación Científica Avanzada, Universidad de Extremadura, 06071 Badajoz, Spain. E-mail: joaquin@unex.es

b) Climate and Atmosphere Research Centre, The Cyprus Institute, Nicosia 2121, Cyprus. E-mail: s.bhowmick@cyi.ac.cy

**Abstract**

Temperature dependence of the thermal rate constants and kinetic isotope effects (KIE) of the CN + $C_2H_6$ gas-phase hydrogen abstraction reaction was theoretically determined within the 25-1000 K temperature range, i.e., from ultra-low to high-temperature regimes. Based on a recently developed full-dimensional analytical potential energy surface fitted to highly accurate explicitly correlated *ab initio* calculations, three different kinetic theories were used: canonical variational transition state theory (CVT), quasiclassical trajectory theory (QCT), and ring polymer molecular dynamics (RPMD) method for the computation of rate constants. We found that the thermal rate constants obtained with the three theories show a V-shaped temperature dependence, with a pronounced minimum near 200 K, qualitatively reproducing the experimental measurements. Among the three methods used in this work, the QCT and RPMD methods have the best agreement with the experiment at low and high temperatures, respectively. The significant increase in the rate constant at ultra-low temperatures in this very exothermic and practically barrierless reaction can be attributed to the large value of the impact parameter, ruling out the role of the tunneling effect and the intermediate complexes in the entrance channel. The theoretical H/D KIE depicted a "normal" behaviour, i.e., values greater than unity, emulating the experimental measurements and previous theoretical results. Finally, the discrepancies between theory and experiments were analysed as a function of several factors, such as limitations of the kinetics theories and the potential energy surface, as well as the uncertainties in the experimental measurements.

# 1. INTRODUCTION

The CN + $C_2H_6$ reaction has drawn considerable interest from field experts in both experimental and theoretical communities as it poses two intriguing challenges, along with some valuable applications. These challenges are associated with i) the potential existence of intermediate complexes in the entrance channel, useful to explain the kinetics models used, and ii) the temperature-dependent plot of rate constant displays a V-shaped pattern, characterized by substantial values at low and high temperatures and markedly smaller values at intermediate temperatures. One of the most interesting applications of the title reaction can be linked to its chemistry occurring at low and ultra-low temperatures, such as in interstellar mediums.

The study of barrierless reactions poses a significant theoretical challenge, particularly for the title reaction, where the complexity is increased due to its polyatomic nature involving 10 bodies and 24 degrees of freedom. We recently developed the first analytical full-dimensional potential energy surface describing the nuclei motion,[1] denoted as PES-2023. These barrierless reactions are typically very fast processes with high exothermicity and proceed with "early" transition states. Therefore, it is imperative to use high-level electronic structure methods for accurate descriptions of such reactions. In general, the rate constants of these reactions exhibit an inverse temperature dependence, i.e., the rate constant increases as the temperature decreases. This is usually a sign of the presence of an intermediate complex in the entrance channel with a "submerged" transition state (TS), i.e., the energy of the TS is lower than the reactants. For more detailed descriptions of such reactions, the reader is referred to the pioneering works of Benson et al.[2] in 1984 and Levine et al.[3] in 1991.

Among other reactions,[4] the reaction between cyano radical and ammonia, recently studied by one of us (JEG),[5] serves as a typical example of a radical-neutral barrierless reaction. This reaction has high exothermicity (releasing 17.5 kcal mol$^{-1}$ of heat), a "submerged" transition state (1.4 kcal mol$^{-1}$ lower energy than the reactants), and a stabilized reactant complex (8.1 kcal mol$^{-1}$ lower energy than the reactants). The reaction is fast, and the corresponding rate constant exhibits an inverse temperature dependence below 300 K. Remarkably when the temperature was lowered from 198 K to 25 K, the rate constant increased by a factor of 17.[6-8] It is, however, worthwhile to note that the presence of an intermediate complex in the entrance channel is not a mandatory condition for the reactions exhibiting an inverse temperature dependence of the rate constants. An example illustrating the above point is the addition reaction between OH and $C_2H_4$.[9] This reaction is also characterized by high exothermicity (-28.9 kcal mol$^{-1}$) and practically zero barrier height. However, no intermediate complex was identified along the reaction pathway. Experimentally,[10-17] a negative temperature dependence was observed for this reaction at temperatures below 500 K, accompanied by a negative activation energy ($\approx$ -1 kcal mol$^{-1}$).

The gas-phase reaction between the cyano radical and hydrocarbons, which is the primary focus of this study, holds significance under different reaction conditions. In high-temperature hydrocarbon fuel combustion processes, for instance, CN plays a crucial role in the formation of nitric oxide.[18,19] Conversely, at low and ultra-low temperatures, it contributes to the chemistry of planetary atmospheres[20] and interstellar clouds.[21,22] The hydrogen abstraction from ethane by CN proceeds through the following mechanism:

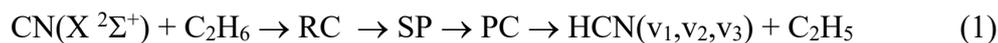

$$CN(X\ ^2\Sigma^+) + C_2H_6 \rightarrow RC \rightarrow SP \rightarrow PC \rightarrow HCN(v_1,v_2,v_3) + C_2H_5 \qquad (1)$$

where RC, SP, and PC are, respectively, the reactant complex, the saddle point, and the product complex.

The abundant kinetic and dynamic experimental measurements on the title reaction (see the complete list in the NIST Chemical Kinetics Database from 1965 to 2013)[23] contrast with the scarce theoretical studies. To the best of our knowledge, only one theoretical study was reported in the literature by Klippenstein et al. in 2007.[24] Kinetically, the title reaction has been studied in a very broad temperature range, from 25 to 1140 K. The rate constant shows an unusual temperature dependence, exhibiting a V-shaped pattern when plotted against temperature. Remarkably high rate constants, in the order of $10^{-10}$ cm$^3$ molecule$^{-1}$ s$^{-1}$, were obtained at both very low (25 K) and very high (1000 K) temperatures, while a minimum of about $10^{-11}$ cm$^3$ molecule$^{-1}$s$^{-1}$ is observed at the intermediate temperatures (~200 K). In order to explain this behaviour, several arguments have been proposed in the literature, such as i) the high electron affinity of the CN radical favours the barrier height for the hydrogen abstraction reaction to be close to zero;[25] ii) the presence of a van der Waals complex in the entrance channel was suggested to explain the hydrogen abstraction mechanism;[26,27] iii) Klippenstein et al.[24] developed a theoretical model based on two transition states (2TS), where at low temperatures, an "outer" TS (close to the reactants) facilitates the reaction, whereas at elevated temperatures, an "inner" TS (close to the barrier) act as a bottleneck for the reaction. Figure 1 depicts the experimentally measured temperature dependence of the rate constants in the wide temperature range of 25-1140 K.[18,26-32] Note that no individual experimental study has measured the rate constants over the whole temperature range. Consequently, the reported rate constants reflect the combinations of different experimental conditions. In general, the hydrogen abstraction reaction is very fast, associated with high exothermicity and very low barrier height.

The primary objective of this work is twofold: (i) to evaluate the role played by the intermediate complex in the entrance channel and (ii) to determine the temperature dependence of the rate constants. The present paper is organized as follows. In Section 2, the PES-2023 surface[1] is briefly described, together with the computational details of three kinetics approaches used in this work: variational transition-state theory with multidimensional tunnelling corrections (VTST/MT), quasi-classical trajectory calculations (QCT), and ring polymer molecular dynamics (RPMD) method. In Section 3, we provide the details of the theoretical results obtained in this work. Firstly, we described the exhaustive set of high-level electronic structure calculations to locate and characterize all stationary points and the reaction path, such as the reactants, products, and saddle points, with special attention given to the intermediate complex in the entrance channel. In addition, the temperature dependence of the rate constant and the kinetic isotope effects (KIEs) were determined by the three different methods and were compared with the available experimental measurements and theoretical reports. Finally, the main conclusions are summarized in Section 4.

## 2. THEORETICAL TOOLS

The quality of the kinetics (and dynamics) results depends on the use of accurate full-dimensional PESs and the employed kinetics theories. To describe the title gas-phase reactive system, we have recently developed a full-dimensional analytical potential energy surface, viz., PES-2023.[1] The PES-2023, which describes the nuclear motion within the Born-Oppenheimer approximation, has been constructed by fitting data obtained from highly accurate explicitly correlated CCSD(T)-F12/aug-cc-pVTZ theory to valence-bond/molecular mechanics (VB/MM) functional form. This PES can describe the topology of this polyatomic reactive system in a smooth and continuous way. For example, it can capture the high exothermicity (-25.55 kcal mol$^{-1}$) very accurately, as evident from the comparison to the experimental data obtained from the corresponding standard enthalpies of formation (-25.6 kcal mol$^{-1}$).[33] Moreover, the low barrier height (+0.23 kcal mol$^{-1}$) and the presence of intermediate complexes in the entrance and exit channels can be identified in PES-2023; the first intermediate, which is moderately stabilised with respect to the reactants (by 0.27 kcal mol$^{-1}$), while the second intermediate has a comparatively larger stabilisation with respect to the products (1.10 kcal mol$^{-1}$). This surface was used in our previous work[1] to perform an exhaustive dynamical study, analysing several dynamical properties, such as the product rotational, vibrational, translational, and angular distributions. The vibrational populations of the three vibrational modes of the product HCN($v_1$,$v_2$,$v_3$), viz., C-N stretch ($v_1$), bend ($v_2$), and C-H stretch ($v_3$), were also determined. We find that the product distributions reasonably reproduced the experimental observations, thereby serving a stringent test for evaluating the quality of the surface.

In the present study, the PES-2023 surface was utilized to explore the temperature dependence of the rate constant of the title reaction over a broad temperature range of 25-1000 K, i.e., from ultra-low to high temperatures, from three different but efficient kinetics approaches, viz., VTST/MT, QCT, and RPMD. In the first method, i.e., VTST/MT, the thermal rate constants were calculated using canonical variational transition-state theory (CVT),[34,35]

$$k^{CVT}(T) = \sigma \frac{k_B T}{h} K^o \min_{s} \exp\left[\frac{-\Delta G^{CVT,o}(T,s^{*,CVT})}{k_B T}\right] \qquad (2)$$

where the usual terms in this theory are: σ represents the symmetry factor or number of equivalent paths (six in this case for the forward reaction). $k_B$, T, and h are, respectively, Boltzmann constant, temperature and Planck constant. $K^o$ represents the reciprocal of the standard-state concentration (1 molecule cm$^3$), and $\Delta G^{CVT,o}$ is the maximum of the free energy of activation associated with the dividing surface $s^{*,CVT}$ along the reaction path. In this statistical theory, the rotational partition functions were calculated classically. The vibrational partition functions were calculated as separable harmonic oscillators using redundant internal coordinates.[36-39] The CVT calculations were performed using the Polyrate-2016 program.[40] Note that since the title reaction is practically barrierless, tunnelling corrections are negligible.

In the second kinetic approach, the thermal rate constants were obtained by QCT theory using the Venus code,[41,42]

$$k(T) = \left(\frac{8k_BT}{\pi\mu}\right)^{1/2} \pi.b_{max}^2 \frac{N_r}{N_T} \tag{3}$$

where μ is the reduced mass, $N_r$ and $N_T$ are, respectively, the number of reactive trajectories and the total number of trajectories run at each temperature. $b_{max}$ is the maximum impact parameter obtained at each temperature by running small batches of trajectories until no reactive trajectories were found. The respective values of $b_{max}$ were decreased from 10.0 Å at 25 K to 5.5 Å at 1000 K. The estimated statistical error (one standard deviation) was given by,[43]

$$\Delta k(T) = k(T)\sqrt{\frac{N_T - N_r}{N_T N_r}} \tag{4}$$

which, given the high reactivity and the great number of trajectories run, is less than 3% and, therefore, will not be reported in the remainder of the paper. The QCT input parameters are summarized in Table 1. Note that the QCT theory is classical in nature, and therefore, quantum effects are not considered. However, since the title reaction is barrierless, the quantum tunnelling effect can be considered negligible. Another point of concern for the QCT theory is that some trajectories end up with vibrational energy below their zero-point energy (ZPE), also known as the ZPE violation problem. To resolve this issue, in the present work, we have considered two approaches: (i) the "All" approach, where all reactive trajectories were considered in the final analysis, and (ii) the "DZPE" (double ZPE approach) approach where only reactive trajectories with the vibrational energy of each product, i.e., HCN and $C_2H_5$, above their respective ZPEs were taken into account. Note that these approaches (as well as others that can be found in the literature) remain intrinsically ad hoc.[44-46]

Finally, the kinetic calculations were also performed using the RPMD method using the RPMDrate package.[47] The RPMD rate constant is obtained as the product of two factors,

$$k_{RPMD}(T) = k_{QTST}(T,\xi^{\ddagger}).\kappa(t \to \infty, \xi^{\ddagger}) \tag{5}$$

The first term is a static factor dependent on temperature and the reaction coordinate $\xi^{\ddagger}$ and is referred to as the centroid-density quantum transition-state theory (QTST) rate constant.[47] This factor has been calculated from the centroid potential of mean force (PMF) along the reaction coordinate. The second factor of equation (5) is a dynamic term called the ring polymer transmission coefficient or ring polymer recrossing factor. This factor ensures that the final rate constant is independent of the choice of the dividing surface along the reaction path, which is an advantage compared to the TST-based approaches. Previous studies have shown that the RPMD method can efficiently take account of the quantum effects, such as the ZPE and tunnelling, along the reaction pathway.[48,49] The RPMDrate input parameters are summarized in Table 2.

## 3. RESULTS AND DISCUSSION

### 3.1. Intermediate complexes

The presence of an intermediate complex in the entrance channel is an unresolved aspect of the title reaction, potentially associated with the negative activation energy. In this section we analyse the various factors which can affect the stability of the intermediate complex using five different theoretical approaches, viz., second order Møller-Plesset perturbation theory (MP2) with the 6-31G (d,p) basis set, MP2/6-31G(d,p); Minnesota functional M06-2X/6-311++G(d,p); Minnesota functional M08-HX/6-311++G(3d2f,2p); coupled cluster method with single, double and triple (perturbative) excitations using the correlation consistent triple zeta basis set, CCSD(T)/cc-pVTZ; and the results obtained from the PES-2023 surface.[1] The geometries and vibrational frequencies were obtained using the Gaussian code[50] for the first four electronic structure calculations. The single-point energies of the geometries obtained from the methods mentioned above are determined using a more accurate explicitly correlated coupled cluster method with a larger basis set, i.e., CCSD(T)-F12/aug-cc-pVTZ, using the MOLPRO code.[51]

Figure 2 shows the geometries of the reactant complex (RC) obtained with the five different methods, and the corresponding vibrational frequencies are reported in Table 3. From Figure 2, it is evident that the geometry of RC obtained from each method and the PES-2023 are different. The geometries of the RC obtained at MP2 and coupled cluster method are cyclic, formed by six atoms (CH-CH of ethane and CN). Although cyclic structures of RC are also obtained at the DFT level, they are five-membered rings (the N atom does not take part). Finally, the CH (of ethane) and CN moieties are almost linear for the RC in PES-2023. This RC complex was characterized as a true minimum at each level of calculation, i.e., all vibrational frequencies were positive. The lowest vibrational frequencies obtained by the different methods, PES-2023 and Ref. 24, have very small values (43-17 cm$^{-1}$), suggesting a very flat surface in the entrance channel. The zero-point energy (ZPE) obtained by each of the methods not only have good agreement among each other but also with Klippenstein et al.[24] (with differences up to ± 1 kcal mol$^{-1}$), except those obtained by the MP2 method. Note that it is known that the MP2 method overestimates the vibrational frequencies. These results highlight the difficulty in locating the reactant complex, which is strongly dependent on the level of calculation.

Table 4 shows the stabilization energies of the RC with respect to the reactants, in which the zero-point energy at 0 K (ZPE) and the thermal corrections (TC) at 298 K were included in order to evaluate the effect of thermal factors on its stability. With the exception of the M08 method, all other methods indicate a modest stabilization of the reactant complex, which falls in the range between -0.27 to -1.11 kcal mol$^{-1}$. However, when the ZPE correction was included, this stabilization diminished. Furthermore, when the thermal correction at 298 K was considered, the RC may not form, given that the corresponding ΔH values were positive. This is true for each of the methods and PES considered in this work. It is interesting to note that these results agree well with the previous theoretical results reported by Klippenstein et al.[24] In that study, the authors employed CASPT2(7e,6o) multi-reference level of calculations with augmented double-zeta basis set and extrapolated the results to the complete basis set limit. Finally, we have also performed a few calculations using a larger quadruple-zeta quality (cc-pVQZ) basis set on the geometries optimized at the CCSD(T)/cc-pVTZ level by employing explicitly correlated CCSD(T) method, i.e., CCSD(T)-F12. We found that the RC is stabilized by -0.94 kcal mol$^{-1}$, which is very similar to the results obtained at the CCSD(T)-F12/aug-cc-pVTZ level of calculations (-1.03 kcal mol$^{-1}$, see Table 4). This agreement indicates that the CCSD(T)-F12/aug-cc-pVTZ level is adequate to represent this intermediate complex with reasonable accuracy. Overall, from the above analyses, it can be stated that the detection of this RC may depend upon factors such as the level of *ab initio* calculations and the inclusion of ZPE and TC corrections, calling into question the existence of this reactant complex. In sum, the small differences found in this work (within the chemical accuracy of ± 1 kcal mol$^{-1}$) also indicate a delicate balance between many factors.

In light of the challenges associated with describing the reactant valley, we have also identified and characterized the saddle point with the same computational methodologies, revealing some unexpected findings. The calculated electronic barrier height (in kcal mol$^{-1}$) and imaginary vibrational frequency (in cm$^{-1}$) for each level are, respectively, 0.31/157 i for MP2; -2.11/58 i for M06; -1.77/138 i for M08; +0.62/196 i for CC/TZ and +0.23/149 i for PES-2023, and those reported by Klippenstein et al.[24] is +0.41/212 i. Note that when the single-point energies are calculated at CCSD(T)-

F12/aug-cc-pVTZ and cc-pVQZ levels on the geometries optimized at the CC/TZ level, the barriers are, respectively, +0.21 and +0.39 kcal mol$^{-1}$. These findings suggest a remarkably low barrier height (negative for the DFT methods) and a low imaginary frequency. This implies a notably flat reactant channel, providing further confirmation of the challenges encountered while attempting to locate and characterize the stationary points, namely the reactant complex and the saddle point. All saddle points were characterized by only one imaginary frequency, i.e., one negative eigenvalue on the reaction coordinate. However, when the eigenvectors were drawn (see Figure 3), we found that for the DFT methods (M06 and M08), the corresponding eigenvectors are associated with umbrella, bending, and torsion motions of the ethane fragment; therefore, they do not represent the hydrogen abstraction reaction. On the other hand, the eigenvectors obtained from the MP2 and CC/TZ methods describe the hydrogen abstraction, although they couple with other motions. Evidently, the clearest description of the hydrogen abstraction reaction was obtained from the PES-2023 (see Figure 3).

## 3.2. Rate constant analysis. Comparison with experiment

Figure 4 illustrates the comparison of rate constants calculated in this work with previous experimental measurements[23] and theoretical reports.[24] In general, the three kinetic approaches considered in this work reproduce the experimental V-shaped temperature dependence of the rate constant. However, the calculated rate constant differs in absolute values. For example, the minimum rate constant was located at 170, 180, and 210 K with the CVT, QCT, and RPMD methods, respectively, reproducing the experimental evidence near 200 K. Considering the broad temperature range investigated in this study (25-1000 K), we separately discuss the rate constants at three different temperature ranges, viz., low (25-250 K), medium (300-700 K) and high (1000 K) temperatures.

We begin by analysing the behaviour of the rate constant at high temperatures where the recrossing effects may play an important role. At 1000 K, the average experimental value[30,31,32] of the rate constant is 0.96($\pm$0.1)$\times$10$^{-10}$ cm$^3$ molecule$^{-1}$ s$^{-1}$. The theoretical approaches, CVT, QCT, and RPMD, give values of 1.21$\times$10$^{-10}$, 0.73$\times$10$^{-10}$ (when all trajectories are considered and 0.64$\times$10$^{-10}$ using the DZPE approach) and 1.02$\times$10$^{-10}$ cm$^3$ molecule$^{-1}$ s$^{-1}$, respectively. The rate constants obtained from the QCT method permit us to evaluate the effect of the two approaches used to consider the ZPE on the kinetics of the title reaction. Both QCT approaches underestimate the experimental measurements by 31% and 50% when all trajectories and DZPE are considered, respectively. In QCT theory, the thermal rate constants are obtained from the corresponding reaction cross sections (see Eq. (3)), the variation of which with temperature is illustrated in Figure 5. We found that both QCT approaches, All and DZPE, display similar behaviour within the overall temperature range considered in this work. More specifically, the reaction cross section decreases monotonically with the increase in temperature, a typical behaviour for non-threshold reactions. Therefore, for an average property like the rate constant, as opposed to the state-to-state properties determined in an earlier dynamical study,[1] the "All" approach should give results comparable to those of the DZPE approach. Therefore, in the remainder of the paper, we discuss the results obtained specifically from the "All" approach of QCT rather than the DZPE approach, given that the latter incorporates additional approximations. We should also note that similar to the QCT results obtained here, the work of Klippenstein et al.[24] also underestimated the experimental measurements at 1000 K ($0.73 \times 10^{-10}$ cm$^3$ molecule$^{-1}$ s$^{-1}$).

The CVT approach overestimates the experimental result by 21%, while the result obtained from the RPMD method has the best agreement with the experiment, with around 5% deviation. Given that the RPMD method is exact at high temperatures (see, for instance, Refs. 47-49 and 52-56), this agreement is indirect evidence of the quality of PES-2023. In the CVT method, the recrossing effect is evaluated as the ratio between CVT (s = s*) and TST (s = 0, saddle point) rate constants, i.e., it measures the effect of the shift of the maxima of the free energy curve, s*, from the saddle point, s = 0, and is commonly referred to as the "variational effect". At 1000 K, this ratio is 0.999, i.e., the recrossing effect is practically negligible, indicating that the maximum of the free energy curve is located at the saddle point. However, the recrossing coefficient obtained from the RPMD method is very small (0.263), indicating high recrossing, as expected in this heavy-light-heavy reaction.[52,56] The above comparison shows that the CVT method strongly underestimates the recrossing effect, and this deficiency is related to the location of the dividing surface between reactants and products. As stated previously, the RPMD method is immune to the choice of the dividing surface.[57] In the CVT approach, the choice of the coordinate system and the harmonic/anharmonic treatment of the vibrational frequencies strongly influence the location of the dividing surface. Finding the optimum dividing surface becomes increasingly intricate for an exceptionally flat entrance channel, such as in the case of the title reaction. Although Isaacson[58,59] proposed an alternative approach to treat the anharmonicity for systems consisting of three and four atoms, to the best of our knowledge, in the case of polyatomic systems composed of five or more atoms, a general procedure for addressing the anharmonicity along the reaction path has not been developed. Finally, it is important to highlight that in the barrierless (or shallow barrier) reactions with very flat reactant channels, for example, F + $NH_3$ or CN + $NH_3$ reactions, significant difficulties in the application of the variational transition state theory related to the location of the dividing surface between reactants and products have been reported previously.[60,61] Therefore, the CVT results reported in the present work must be taken as a first approximation.

Next, we analyse the opposite situation of the low-temperature regime (25-250 K), where the rate constant increases monotonically with the decrease in temperature. It is important to highlight that since the title reaction is practically barrierless, quantum mechanical tunnelling correction to the rate constant is expected to be negligible. The experimental data at the low-temperature regime is very limited, with only one such report available in the literature.[26] In comparison to the experimental measurement and the theoretical calculations reported by Klippenstein et al.,[24] in this work, the best agreement was obtained for the QCT (All approach) method with an average error of 13 %. On the other hand, the CVT approach underestimates the experimental rate constants (with a 61 % error), while the RPMD method slightly overestimates it (with a 34 % error). The good agreement obtained with the QCT approach should not be generalized to other reactions since, in principle, it is classical in nature. The QCT method for title reaction is particularly a good approximation due to the fact that the tunnelling effect is negligible. Examining the evolution of the maximum impact parameters, denoted as $b_{max}$ in Eq. (3), reveals a substantial increase in this parameter as temperature decreases, rising from 5.5 to 10.0 Å for the lowering in temperature from 1000 K to 25 K. Based on this, we propose that the significant increase in the rate constant at low temperatures can be attributed to this parameter. This increase in the maximum impact parameter values is a typical feature of barrierless reactions and has been reported in other similar reactions, for instance, S + OH or Si + OH.[62,63] The overestimation observed in the RPMD rate constants (on average, a factor of 1.5 within this temperature range) reflects the overall trend identified with this theory at low temperatures, particularly in the case of asymmetric reactions.[64,65] Nevertheless, the overestimation in the rate constant of the title reaction is markedly smaller than those obtained previously for other reactions, where factors of 2-3 were reported.[52-54, 66-68] This smaller error for the title reaction may be attributed to the absence of tunnelling effects.

At intermediate temperatures, i.e., within 300-500 K, the three kinetics approaches are in good agreement with the experiment. For instance, at 500 K, the CVT, QCT and RPMD rate constants are, respectively, $3.83\times10^{-11}$, $3.65\times10^{-11}$ and $4.98\times10^{-11}$ cm$^3$ molecule$^{-1}$ s$^{-1}$. These values align well with the average experimental value of five different studies[18, 28-31] having small deviations among themselves $(3.75\pm0.10)\times10^{-11}$ cm$^3$ molecule$^{-1}$ s$^{-1}$ or with the theoretical result of Klippenstein et al.[24] ($3.89\times10^{-11}$ cm$^3$ molecule$^{-1}$ s$^{-1}$).

Examining the values of phenomenological activation energy also allows us to ascertain the accuracies of theoretical tools, such as kinetics theories and the underlying PES. The activation energies ($E_a$) were obtained from the slopes of the plots of the rate constants with temperature (see Figure 4). For instance, the values of $E_a$ determined from the CVT method are as follows: (i) -0.25 kcal mol$^{-1}$ (within the 25-250 K range), (ii) +1.03 kcal mol$^{-1}$ (within the 300-500 K range), and (iii) +3.39 kcal mol$^{-1}$ (within the 700-1000 K range), where similar values were obtained with the other theories. This temperature dependence of the activation energy can be explained from a thermochemical kinetic analysis[69] within the framework of the transition state theory. The activation energy can be expressed as a function of the enthalpy change, $\Delta H(T)$, and the temperature as,

$$E_a = \Delta H(T) + 2RT, \qquad (6)$$

where R is the universal gas constant. At the transition state, the enthalpy change has a small negative value (-0.38 kcal mol$^{-1}$), and consequently, the positive or negative value of $E_a$ is a delicate balance between enthalpy and temperature. Therefore, at very low temperatures, the enthalpic term dominates, resulting in negative activation energy. On the other hand, at elevated temperatures, the dominating factor is the 2RT term, leading to positive $E_a$ values.

### 3.3. Kinetic isotope effects

The kinetic isotope effects (KIEs) serve as a powerful tool for gaining insights into the reaction mechanism and assessing the influence of factors such as the recrossing and tunnelling (negligible for the title reaction) on reactivity. The KIEs are defined as the ratio between the rate constants of the lighter to the heavier isotope, R1/R2,

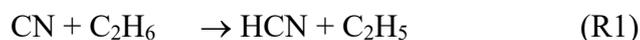
$$CN + C_2H_6 \rightarrow HCN + C_2H_5 \qquad (R1)$$

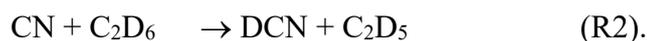

$$CN + C_2D_6 \rightarrow DCN + C_2D_5 \qquad (R2).$$

KIEs are less sensitive to the accuracy of the PES and methods since the errors associated with the kinetic models are essentially nullified. There is only a single set of experimental[29] and theoretical[24] data available for the KIEs concerning the title reaction, and they cover the temperature range of 294-736 K. Figure 6 depicts the KIEs, i.e., R1/R2, obtained in the present study for both CVT and QCT approaches, alongside available experimental and theoretical results. Note that the KIEs for the RPMD method were not calculated, primarily due to the exceedingly high computational cost associated with such simulations. Computed KIEs with the CVT and QCT methods show that they decrease with the increase in temperature, corroborating experimental observations. Moreover, the KIEs reported in this study have better agreements with the experiment than those calculated previously.[24] Within the considered temperature range, the KIEs exhibit a "normal" behaviour, i.e., they are larger than unity. Given that the tunnelling effect is negligible for the title reaction, KIE values can be directly related to the adiabatic barrier height, i.e., classical barrier height and the ZPE corrections. While the classical barrier is the same for both isotopes (+0.23 kcal mol$^{-1}$), the lighter H atom has a lower adiabatic barrier than the heavier D isotope, -0.36 and -0.12 kcal mol$^{-1}$, respectively.

## 4. Conclusions

Based on an accurate analytical full-dimensional PES (PES-2023), in this work, we have performed an exhaustive theoretical kinetic analysis on the CN + CH$_3$CH$_3$ → HCN+ CH$_3$CH$_2$ gas-phase reaction within a wide temperature range of 25–1000 K, i.e., from ultra-low to the combustion regimes. We have used three different kinetic approaches, viz. CVT, QCT, and RPMD to compute the rate constants. The theoretical results were compared with the available experimental measurements. However, caution is warranted in this comparison, given that both theoretical tools, PES-2023 and kinetics approaches, are being tested simultaneously, as well as the presence of experimental uncertainties.

The investigated reaction is highly exothermic (-25.55 kcal mol$^{-1}$), with a very shallow classical barrier (of height +0.23 kcal mol$^{-1}$). The presence of intermediate complexes in the entrance channel of this reaction remains unresolved. Indeed, through different electronic structure theories, such as DFT, MP2, and CCSD(T), including ZPE and thermal corrections in this work, we found that the existence of such a reactant complex is rather uncertain. This reactant complex was previously employed in the literature to account for the temperature dependence of the rate constants.

Experimentally, the rate constants exhibit a V-shaped temperature dependence within the wide temperature range, displaying a pronounced minimum near 200 K. This distinctive characteristic of the rate constants was reproduced in the present work by the three kinetic approaches, all based on PES-2023, thereby serving as a rigorous test of the accuracy of this surface. Within the low-temperature regime (25-250 K), the best agreement with the experiment was obtained for the QCT approach with a 13% difference, while the RPMD and CVT theory systematically overestimates and underestimates the experimental rate constants by 34% and 61%, respectively. At high temperatures (1000 K), where the recrossing effects play an important role, the RPMD theory is accurate, thus allowing us to minimize the theory/experiment differences in the rate constant value associated with the PES to about 5%. At low and ultra-low temperatures, the rate constants increase with the decrease in temperature. This increase is not due to the tunneling effects or the presence of intermediate complexes in the entrance channel but can be attributed to the substantial increase in the impact parameter in QCT theory within this temperature regime.

In summary, reasonable agreement in the calculated rate constants and kinetic isotope effects was found with the experiment for the title reaction, which is difficult to describe due to the presence of a very low barrier and very flat entrance channel, thus demonstrating the efficiency of the PES-2023 employed in this study.

**ACKNOWLEDGEMENTS**

JEG gratefully acknowledges the computer resources at Lusitania (COMPUTAEX) and technical support provided by COMPUTAEX. SB acknowledges the financial support of the Republic of Cyprus through the Research and Innovation Foundation project ML-NANOCAT (CODEVELOP- GT/0322/0093). This research was also supported by the EMME-CARE project, which has received funding from the European Union's Horizon 2020 Research and Innovation Program under Grant Agreement No. 856612, as well as co-funding by the Government of the Republic of Cyprus. SB thanks the Milan High-Performance Computing Facility of The Cyprus Institute for its computational resources.

**CONFLICTS OF INTEREST**

There are no conflicts to declare.

**DATA AVAILABILITY STATEMENT**

The data that support the findings of this study are available from the corresponding author upon reasonable request.

**Table 1.** Input parameters for the quasi-classical trajectory (QCT) calculations[a] for the CN + CH$_3$CH$_3$ reaction

| Parameter | CN+ CH$_3$CH$_3$ | Explanation |
|---|---|---|
| Temperature, T | 25, 50, 100, 150, 200, 250, 300, 500, 700, 1000 | Temperature (K) |
| b$_{max}$ | 10.0, 7.5, 7.1, 7.0, 6.7, 6.6, 6.5, 6.4, 6.0, 5.5 | Maximum impact parameter (Å) |
| Trajectories per T | 1 00 000 | Number of trajectories run |
| Reactant vibration | Thermal sampling | CN and CH$_3$CH$_3$ vibrational energy at each temperature |
| Reactant rotation | Thermal sampling | CN and CH$_3$CH$_3$ rotational energy at each temperature |
| CN-C distance | 15.0 | Initial and final CN-C separation (Å) |
| ε | 0.1 | Propagation step (fs) |
| Impact parameter, vibrational phases and spatial orientation | Monte Carlo sampling | |

a) A more complete explanation of these parameters can be found in the VENUS code manual.[41,42]

**Table 2.** Input parameters for the RPMDrate calculations[a] on the CN + CH$_3$CH$_3$ reaction.

| Parameter | Reaction CN + C$_2$H$_6$ | Explanation |
|---|---|---|
| Command Line Parameters | | |
| T | 25, 50, 100, 150, 200, 250, 300, 500, 1000 | Temperature (K) |
| $N_{beads}$ | 128(25-250K), 64 (300K), 32(500K), 16 (1000K) | Number of beads |
| Dividing Surface Parameters | | |
| $R_\infty$ | 15 a.u. | dividing surface parameter (distance) |
| $N_{bond}$ | 1 | number of forming and breaking bonds |
| $N_{channel}$ | 6 | number of equivalent product channels |
| thermostat | "Andersen" | thermostat option |
| Biased Sampling Parameters | | |
| $N_{windows}$ | 121 | number of windows |
| $\xi_1$ | -0.05 | center of the first window |
| $d\xi$ | 0.01 | window spacing step |
| $\xi_N$ | 1.15 | center of the last window |
| $dt$ | 0.0001 | time step (ps) |
| $k_i$ | 2.72 | umbrella force constant ($(T/K)$ eV) |
| $N_{trajectory}$ | 100 | number of trajectories |
| $t_{equilibration}$ | 20 | equilibration period (ps) |
| $t_{sampling}$ | 100 | sampling period in each trajectory (ps) |
| Potential of Mean Force Calculation | | |
| $\xi_0$ | -0.05 | start of umbrella integration |
| $\xi$ | 1.14 | end of umbrella integration |
| $N_{bins}$ | 5000 | number of bins |
| Recrossing Factor Calculation | | |
| $dt$ | 0.0001 | time step (ps) |
| $t_{equilibration}$ | 20 | equilibration period (ps) in the constrained (parent) trajectory |
| $N_{totalchild}$ | 100000 | total number of unconstrained (child) trajectories |
| $t_{childsampling}$ | 2 | sampling increment along the parent trajectory (ps) |
| $N_{child}$ | 100 | number of child trajectories per one initially constrained configuration |
| $t_{child}$ | 0.5/5.0 (25 K) | length of child trajectories (ps) |

[a] The explanation of the format of the input file can be found in the RPMDrate code manual (https://greengroup.mit.edu/rpmdrate)

**Table 3.** Vibrational frequencies of the reactant complex obtained from different levels of electronic structure calculations: MP2 ≡ MP2/6-31G(d,p); M06 ≡ M062X/6-311++G(d,p); M08 ≡ M08HX/6-311++G(3d2f,2p); CC/TZ ≡ CCSD(T)/cc-pVTZ, and PES-2023. Frequencies are given in cm$^{-1}$. Zero-point energy (ZPE) values (in kcal mol$^{-1}$) are provided at the bottom.

| MP2 | M06 | M08 | CC/TZ | PES-2023 | Ref. 24 |
|---|---|---|---|---|---|
| 3238 | 3092 | 3112 | 3121 | 3022 | 3164 |
| 3235 | 3087 | 3112 | 3118 | 3018 | 3160 |
| 3214 | 3066 | 3080 | 3097 | 3007 | 3140 |
| 3207 | 3055 | 3056 | 3094 | 3004 | 3138 |
| 3138 | 3006 | 3034 | 3041 | 3000 | 3065 |
| 3124 | 2995 | 3004 | 3034 | 2984 | 3060 |
| 2866 | 2188 | 2238 | 2114 | 2114 | 2002 |
| 1571 | 1475 | 1498 | 1515 | 1534 | 1497 |
| 1568 | 1475 | 1494 | 1515 | 1531 | 1494 |
| 1564 | 1468 | 1490 | 1512 | 1510 | 1493 |
| 1564 | 1458 | 1490 | 1512 | 1491 | 1490 |
| 1485 | 1404 | 1415 | 1427 | 1430 | 1411 |
| 1453 | 1382 | 1391 | 1405 | 1393 | 1384 |
| 1269 | 1196 | 1207 | 1226 | 1175 | 1212 |
| 1268 | 1196 | 1205 | 1225 | 1009 | 1210 |
| 1048 | 1020 | 1015 | 1013 | 1002 | 1028 |
| 848 | 802 | 808 | 825 | 828 | 815 |
| 844 | 798 | 808 | 821 | 823 | 812 |
| 343 | 323 | 309 | 321 | 289 | 325 |
| 83 | 126 | 102 | 92 | 90 | 77 |
| 76 | 84 | 65 | 74 | 85 | 65 |
| 62 | 62 | 57 | 57 | 42 | 64 |
| 38 | 52 | 48 | 32 | 33 | 25 |
| 30 | 43 | 38 | 18 | 32 | 17 |
| **ZPE** | | | | | |
| 53.10 | 49.82 | 50.14 | 50.33 | 49.24 | 50.24 |

**Table 4.** Changes in the energy (ΔE) and enthalpies (ΔH) at 0 K and 298 K (in kcal mol$^{-1}$) of the reactant complex (RC) with respect to the reactants obtained at different levels of electronic structure calculations. Except for the PES-2023 and Ref. 24, these values were obtained from single point energy calculations at the CCSD(T)-F12/aug-cc-pVTZ level, based on the geometries optimized at each level of calculation.

|  | MP2 | M06 | M08 | CC/TZ | PES-2023 | Ref. 24 |
|---|---|---|---|---|---|---|
| **ΔE** | -1.11 | -0.50 | +0.24 | -1.03 | -0.27 | -0.91 |
| **ΔH(0K)** | -0.72 | -0.15 | +0.48 | -0.62 | -0.09 | +0.28 |
| **ΔH(298K)** | +0.37 | +0.84 | +0.51 | +0.47 | +0.41 | na |

**FIGURE CAPTIONS**

**Figure 1.** Experimental rate constant, $k$, (in $10^{-11}$ cm$^3$ molecules$^{-1}$ s$^{-1}$) for the CN + C$_2$H$_6$ reaction within the temperature range of 25 to 1140 K.[18, 26-32] The black dashed line is used as a guide.

**Figure 2.** Optimized geometries of the reactant complex using different levels of electronic structure calculation: MP2 ≡ MP2/6-31G(d,p); M06 ≡ M062X/6-311++G(d,p); M08 ≡ M08HX/6-311++G(3d2f,2p); CC/TZ ≡ CCSD(T)/cc-pVTZ, and PES-2023. Distances are given in Å.

**Figure 3.** Optimized geometries and eigenvectors at the saddle point using different levels of electronic structure calculation (see Figure 2 caption for employed methodologies).

**Figure 4.** Variation of the rate constants with temperature obtained in this work using canonical variational transition-state theory (CVT), quasi-classical trajectory (QCT), and ring polymer molecular dynamics (RPMD) methods. Klipenstein-2007 refers to the rate constant values reported by Klippenstein et al.[24] The black dashed line represents the average value from the experimental measurements (see Fig. 1) and is used only as a guide.

**Figure 5.** Reaction cross section (in Å$^2$) with statistical errors < 3% as a function of temperature (in K), obtained using all reactive trajectories (solid line) and reactive trajectories with the double zero-point energy (DZPE) constraints (dashed line).

**Figure 6.** H/D kinetic isotope effects (KIEs) as a function of the temperature (K) within the temperature range 294−736 K, using canonical variational transition-state theory, CVT, (blue line) and quasi-classical trajectory, QCT, (brown line) methods on the PES-2023 surface.[1] Experimental values (black line) are taken from ref 29, and theoretical results (dashed black line) are from ref. 24.

**FIGURE 1**

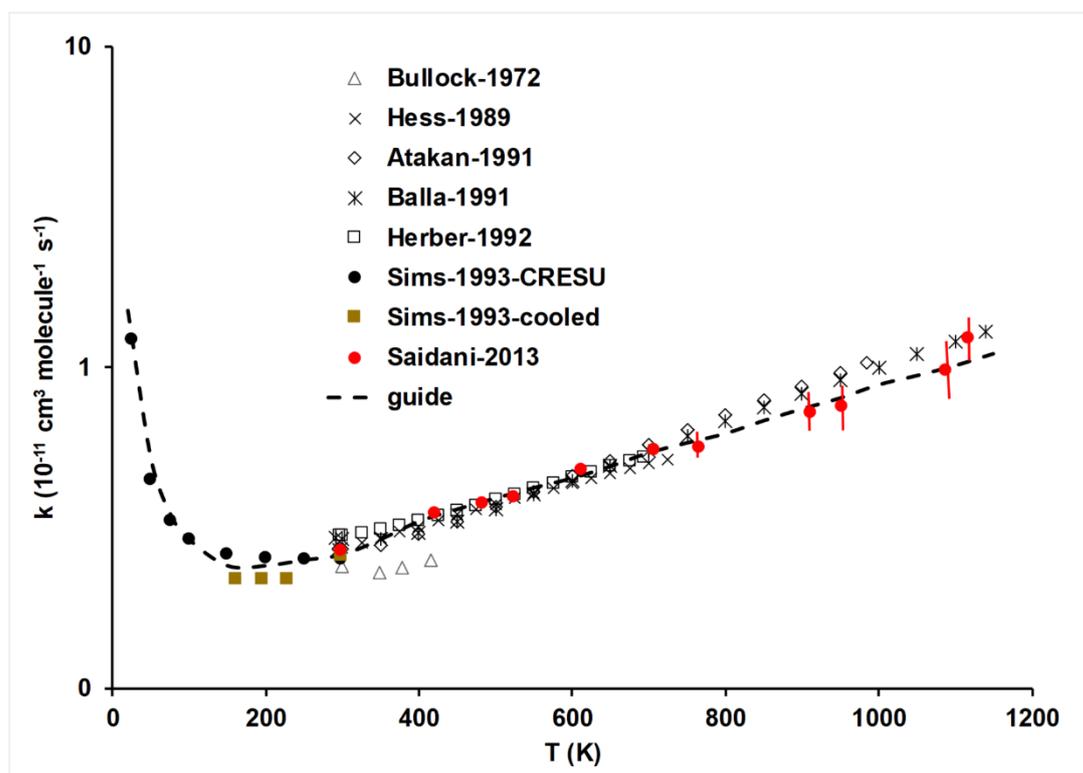

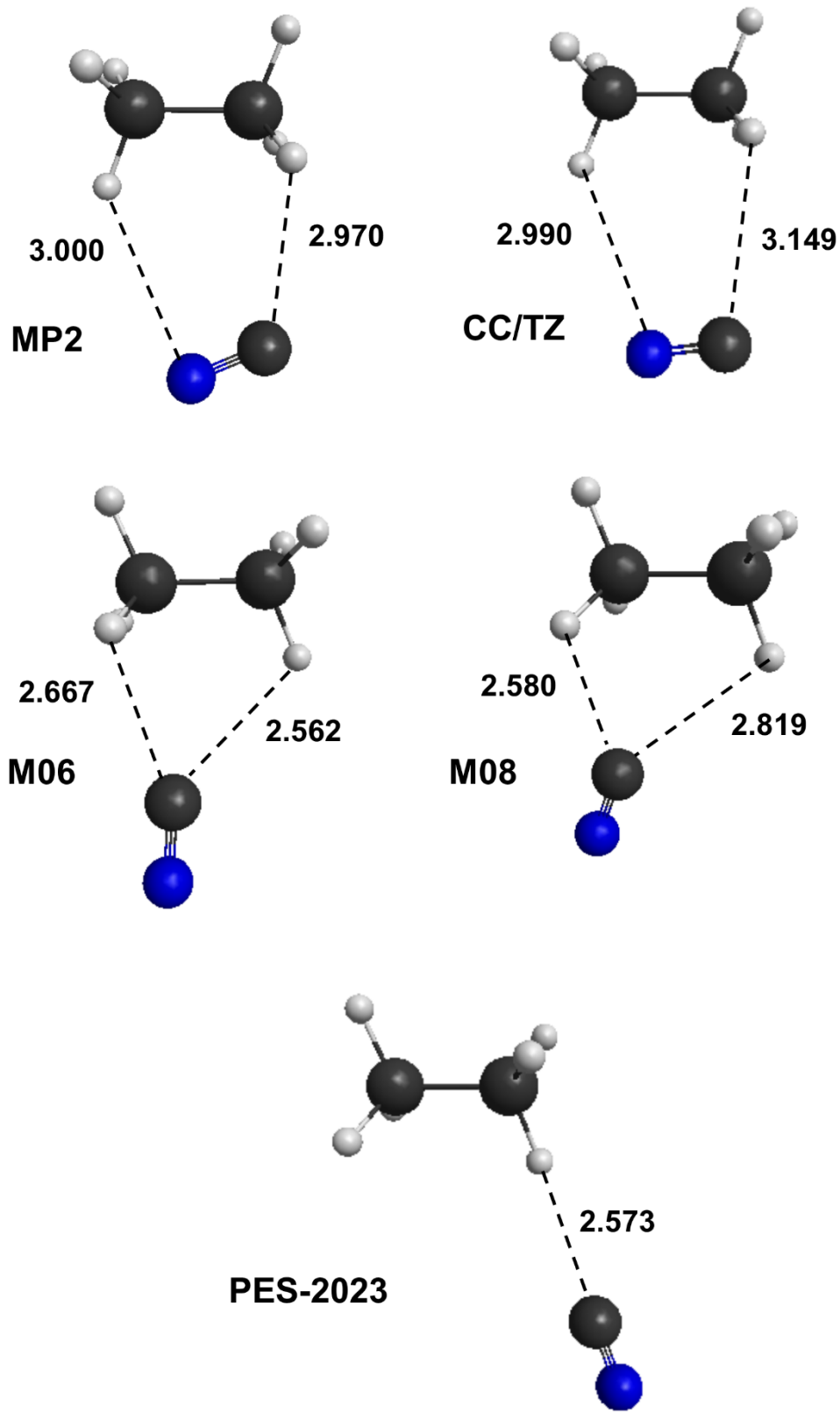

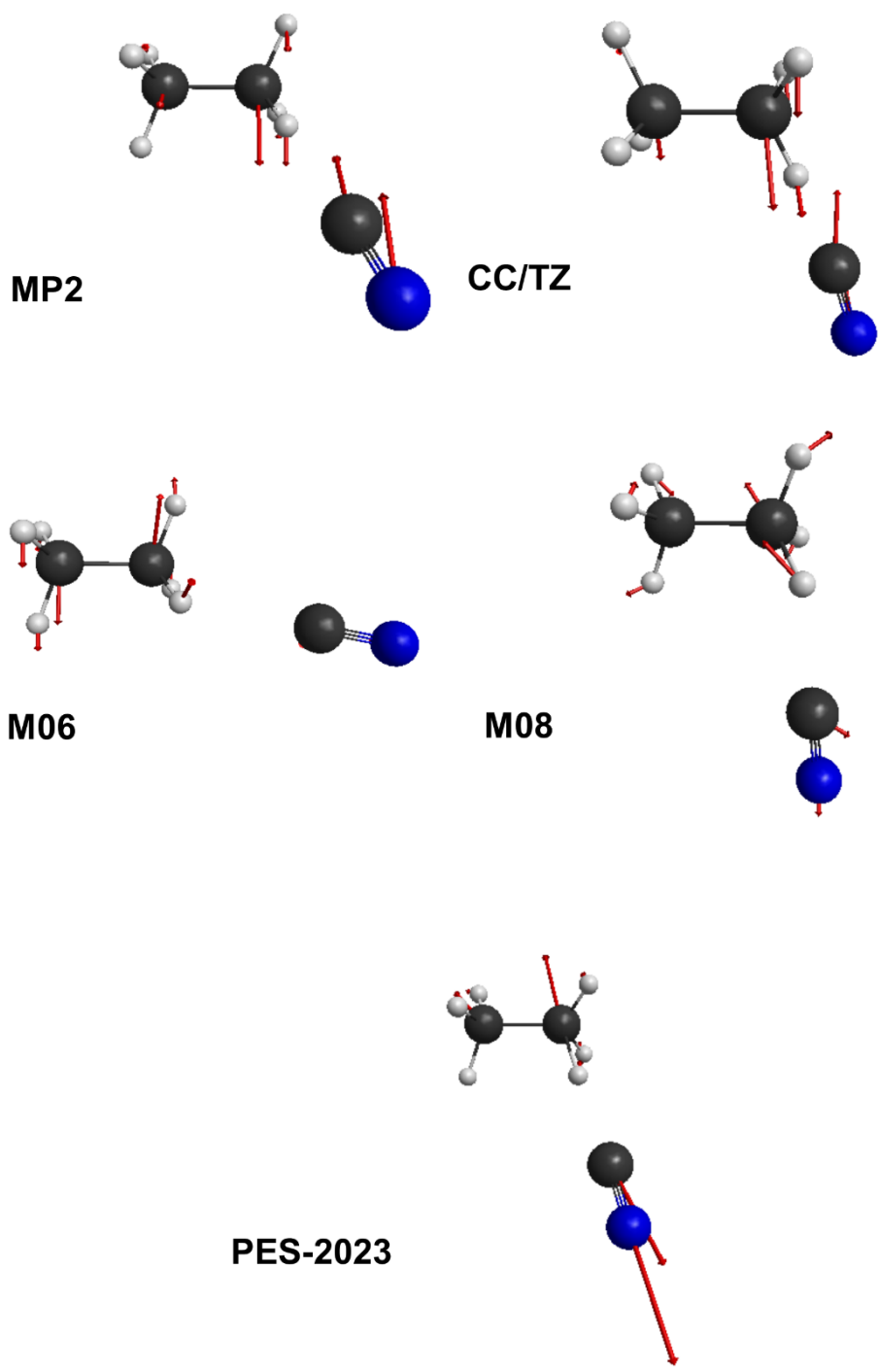

**FIGURE 4**

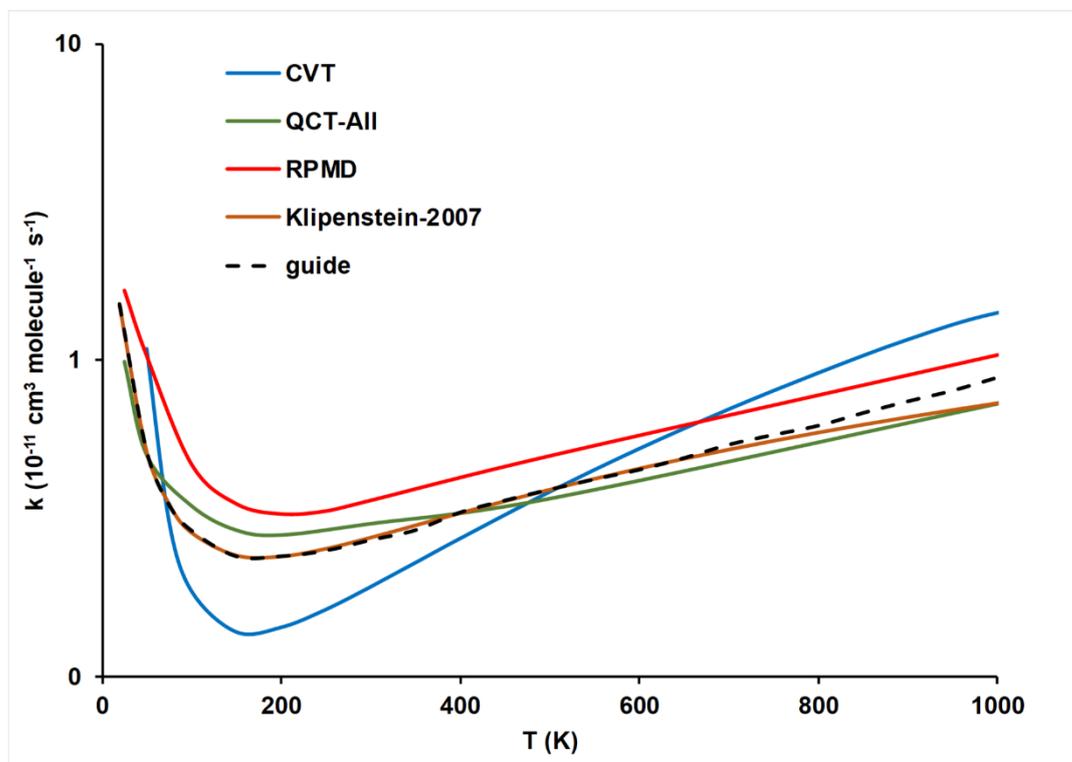

**FIGURE 5**

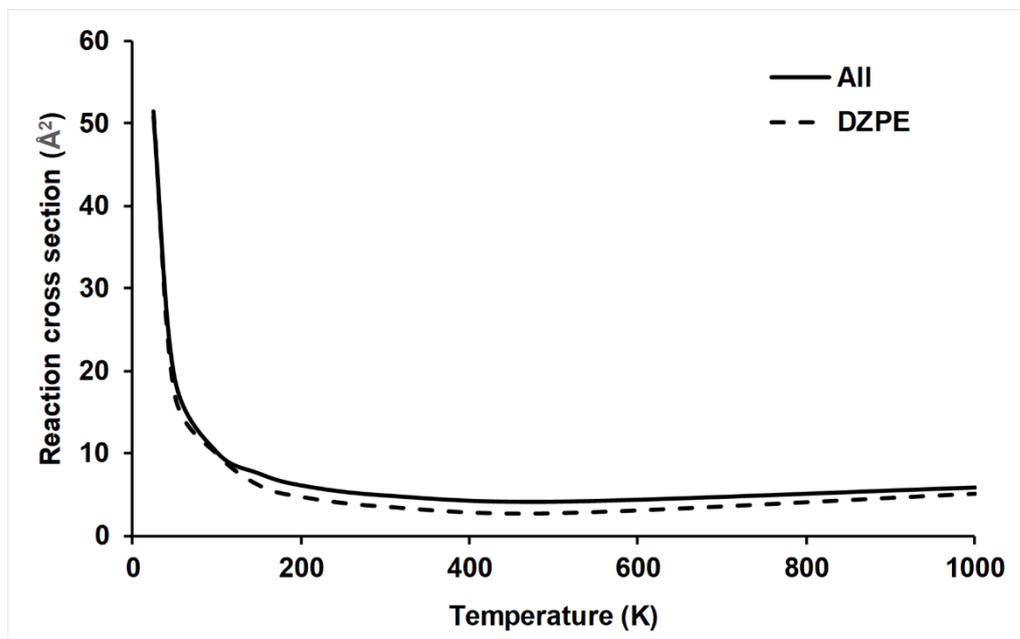

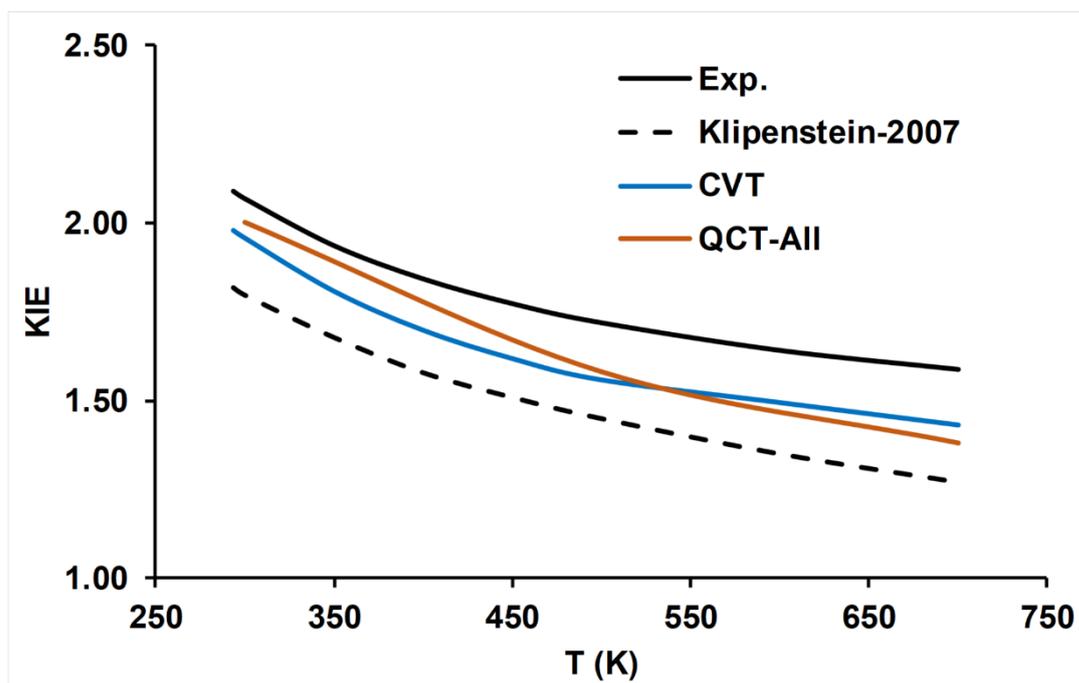